\documentclass[journal]{IEEEtran}
\usepackage{hyperref}
\usepackage{titling} 

\usepackage[natbibapa]{apacite} 
\usepackage{amsmath}

\usepackage{pdfpages}
\usepackage{url}
\usepackage{graphicx}  
\usepackage{float}  

\begin{document}

\title{\ Strategies in JPEG compression using Convolutional Neural Network(CNN)\\ }

\author{Suman Kunwar\\
    \textit{Asia Pacific University of Technology \& Innovation (APU)}\\
    \href{mailto:TP066707@mail.apu.edu.my}{\textcolor{blue}{TP066707@mail.apu.edu.my}}\\}
     
\date{\today}
        

\maketitle

\begin{abstract}
Interests in digital image processing are growing enormously in recent decades. As a result, different data compression techniques have been proposed which are concerned mostly in minimization of information used for the representation of images. With the advances of deep neural networks, image compression can be achieved to a higher degree. This paper describes an overview of JPEG Compression, Discrete Fourier Transform (DFT), Convolutional Neural Network (CNN), quality metrics to measure the performance of image compression, and discuss the advancement of deep learning for image compression mostly focused on JPEG, and suggests that adaptation of model improve the compression.
\end{abstract}


\section{Introduction}

\IEEEPARstart{J}{peg} is one of the most commonly used compression techniques that has been widely
accepted as a standard for lossy image compression. There are two types of image
compression techniques: lossy and lossless compression 
\citep{wang2021a}. Lossy image compression
techniques are non-reversible and can achieve a higher compression ratio whereas
lossless methods provide the best visual experience. Lossy compression comes with a
trade-off between file size and decomposed image quality. In practice, lossy
compression schemes are often preferred on consumer devices because of their much
higher compression ratio \citep{wang2021a}.
\newline
Convolutional Neural Networks are similar to Neural Networks. It consists of neurons
that have learnable weights and biases. Each neuron receives some input, performs a
dot product, and optionally follows it with a non-linearity \citep{kunwar_jpeg_2018}. Deep Convolutional Neural
Networks (ConvNets) is one of the essential tools for computer vision, used in image
classification \citep{ren_faster_2016}, object recognition \citep{7298965}, and semantic segmentation  \citep{cavigelli_accelerating_2015}. It has also
gained relevance for regression tasks in low-level image and video processing by
computing saliency maps  \citep{dosovitskiy_flownet:_2015}, optical flow fields \citep{10.1007/978-3-319-10593-2_13}, and single-frame super-resolution
images \citep{poor_introduction_1988} with state-of-the-art performance.
\newline
This paper begins with the discussion of JPEG compression, Discrete Fourier Transform(DFT) and Convolutional Neural Networks (CNN). Then introduces the concept of the JPEG compression technique and  Convolutional Neural Networks (CNN), and discusses some quality metrics used in image compression. Later talk about the advancement that has happened in JPEG compression using CNN. Finally, a conclusion section ends the chapter.

\section{JPEG Compression}
JPEG has different standards to compress an image such as JPEG, JPEG-LS, JPEG-2000. Two basic image compression algorithms \citep{wallace_jpeg_1991} has been developed by the Joint Photographic Experts Group (JPEG), one of which defines a combination of prediction method and entropy coding, and the other defines a hybrid compression method based on discrete cosine transform (DCT) \cite{rao_discrete_2019} and entropy coding. The former method is a lossless compression technique based on the Differential Pulse Code Modulation (DPCM) and later is a lossy compression technique and is used mostly cause of the high compression ratio.
 
\begin{figure}[H]
\begin {center}
\includegraphics[scale=0.6]{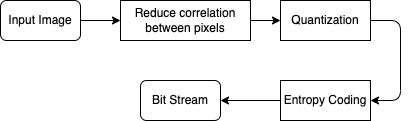}
\caption{Constitution of image coding algorithm}
\label{image coding algorithm}
\end {center}
\end{figure}

Image coding algorithm focuses on reducing the correlation exists between pixels, quantization and entropy coding  \citep{kunwar_image_2017}.
Constitution of image coding algorithm is shown in \autoref{image coding algorithm}. 

\subsection{Source Encoder}
The source encoder aims to decorrelate the input signal by transforming its representation. where the set of data values is sparse, thereby condensing the information content of the signal into the smaller number of coefficients.
\subsection{Quantizer}
A quantizer aims to reduce the number of bits needed to store the transformed coefficients by reducing the precision of these values. Quantization is performed on a per coefficient basis, i.e., Scalar Quantization (SQ) or
Vector Quantization (VQ).
\subsection{Entropy Coding}
Entropy coding aims to eliminate redundancy by removing repeated bit patterns. The most common entropy coders are Huffman coding, arithmetic coding, run-length coding (RLE), and the Lempel-Ziv
(LZ) algorithm.
\newline
\\The basic process of JPEG compression along with entropy encoder is defined in \autoref{jpeg_and_entropy}.
The coding process can be summarized into four parts as shown in red font in Fig. \autoref{jpeg_and_entropy}(a), namely uniform partitioning, transform encoding (FDCT), quantizer, and entropy encoder (\autoref{jpeg_and_entropy}(b)).

\begin{figure}[H]
\begin {center}
\includegraphics[width=0.48\textwidth]{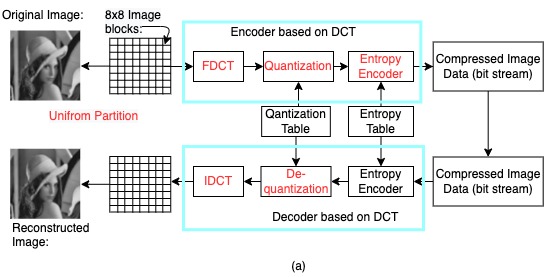}
\label{jpeg_and_entropy}
\end {center}
\end{figure}

\begin{figure}[H]
\begin {center}
\includegraphics[width=0.48\textwidth]{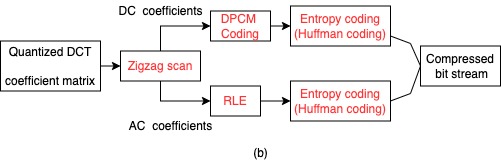}
\caption{Block diagrams of JPEG and Entropy \citep{9088290}}
\label{jpeg_and_entropy}
\end {center}
\end{figure}

 The entropy coder consists of the Zig-Zag scan, Differential Pulse Code Modulation (DPCM) coding, run-length encoding (RLE) followed by Huffman coding. 
 Zig-Zag scan groups lower frequency coefficients in top of vector and high frequency coefficients at the bottom, and maps 8 x 8 matrix to a 1 x 64 vector. Differential Pulse Code Modulation coding encode the
difference between the current and previous 8x8 block. Here, smaller number represents the fewer bits.
The resultant vector obtain from DPCM contains lots of zeros as a result of a quantization table or due to higher entries in the vector capture higher frequency (DCT) where components tend to capture less of the content. RLE encodes series of zeros as skip value pair, where skip denotes number of zeros and value is the next non-zero component.  
 Huffman coding follows the rule of assigning a shorter code to the most frequently used words and a longer code to the less frequently used words.
 Decoding is performed in the reverse order of encoding (i.e., entropy decoder, de-quantizer, and IDCT).
 
\section{Discrete Fourier transform (DFT)}
  Signal and Image processing is one of the emerging technology that touches most of the fields of engineering. Digital Signal signal is used in  applications like data compression, filtering, image processing power spectral estimation, adaptive filtering,  multirate system design and many more. The applications of the discrete Fourier transform are numerous once the fast Fourier transform (FFT) algorithms become available \citep{ifeachor_digital_2002}. Using the symmetry property, data compression is effectively performed by storing only half of the DFT response. To reconstruct the signal, the complete response is required. This can be achieved by considering the complex conjugates of the previous DFT values according to the property and can be checked with Matlab using some non-real-time data \citep{lim_advanced_1988}.
 
 DTFT response is continuous, making it difficult to perform calculations with a digital computer. Because of the infinite spectrum, this is even more difficult during multiplication. To make the computation easier, Discrete Fourier Transform (DFT) is introduced. In DFT, the sampling is done in the frequency domain, i.e. discrete, the corresponding data sequence must be periodic in the time domain \cite{mitra_digital_2006}. 
 \newline
 Given $N$ real numbers
\[
    f_0, \ldots, f_{N-1}
\]
\enspace \enspace \enspace we have to compute 
\begin{equation}   \label{eq:dft-def}
    c_k =  \frac{1}{N} \sum_{j=0}^{N-1} f_j e^{-ijk 2\pi/N }, \quad k = 0, \ldots, N-1.
\end{equation}
\newline
The mapping of $f_j$, $j = 0, \ldots, N-1$, into
$c_k = \frac{1}{N} \sum_{j=0}^{N-1} f_j e^{-ijk 2\pi/N}$, $k = 0, \ldots, N-1$,
is called the {\em Discrete Fourier Transform} (DFT).  {\em inverse discrete Fourier transform} maps $c$ into $f$ and
is given by

\begin{equation}   \label{eq:idft-def}
    f_k =  \sum_{j=0}^{N-1} c_j e^{ijk 2\pi/N }, \quad k = 0, \ldots, N-1.
\end{equation}

\begin{figure}[H]
\begin {center}
\includegraphics[width=0.5\textwidth, height=4cm]{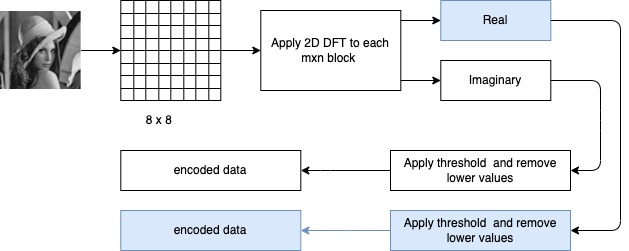}
\caption{JPEG image compression through DFT}
\label{dft_archt}
\end {center}
\end{figure}

\noindent In a straight forward implementation, the computations of matrix-vector products require $N^2$ multiplications.
 JPEG compression in JPEG through is shown in \autoref{dft_archt}

 In DFT transformation, DFT transforms the image symbols into certain coefficients, making it possible to compress by setting a threshold. These thresholds are applied later to the whole image or a certain part of the image part, usually in the form of squares. Here, irreverent information or less important information is discarded. The selection of the size of the block where it will be applied is also relevant to the quality of the image \citep{pratt2006a}.  
 
\begin{figure}[H]
\begin {center}
\hspace*{-1cm}\includegraphics[width=0.6\textwidth, height=4cm]{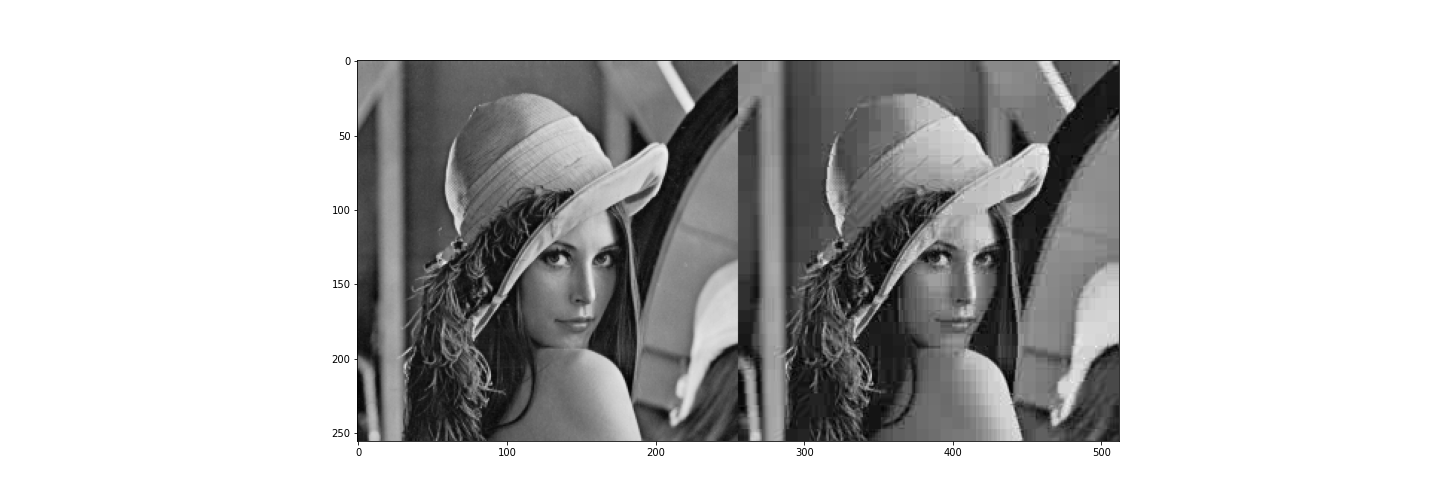}
\caption{Original image \& Compressed image}
\label{meanstd}
\end {center}
\end{figure}

\section{Convolutional Neural Networks (CNN)}
Usage of Convolutional Neural Networks (CNNs) for classification and computer vision tasks are gradually increasing, and have achieved great success in image recognition\citep{krizhevsky_imagenet_2017}. 
It provides a more scalable approach to object detection and image classification tasks.
Supervised Learning and Unsupervised Learning are the two main learning  paradigms in image processing tasks \citep{oshea2015introduction}. In supervised learning pre-labeled inputs act as targets.

Supervised learning is best suited for those problems which has a reference point to train the algorithm. The expected result is known upfront. Unsupervised learning, in contrast, does not contain labels in the training set. Algorithms are left on their own to discover and display the interesting structures in the data. CNN consists of three layers: \textbf{convolutional layers}, \textbf{pooling layers}, and \textbf{fully connected layers}.

\begin{figure}[H]
\begin {center}
\hspace*{-1cm}\includegraphics[width=0.56\textwidth]{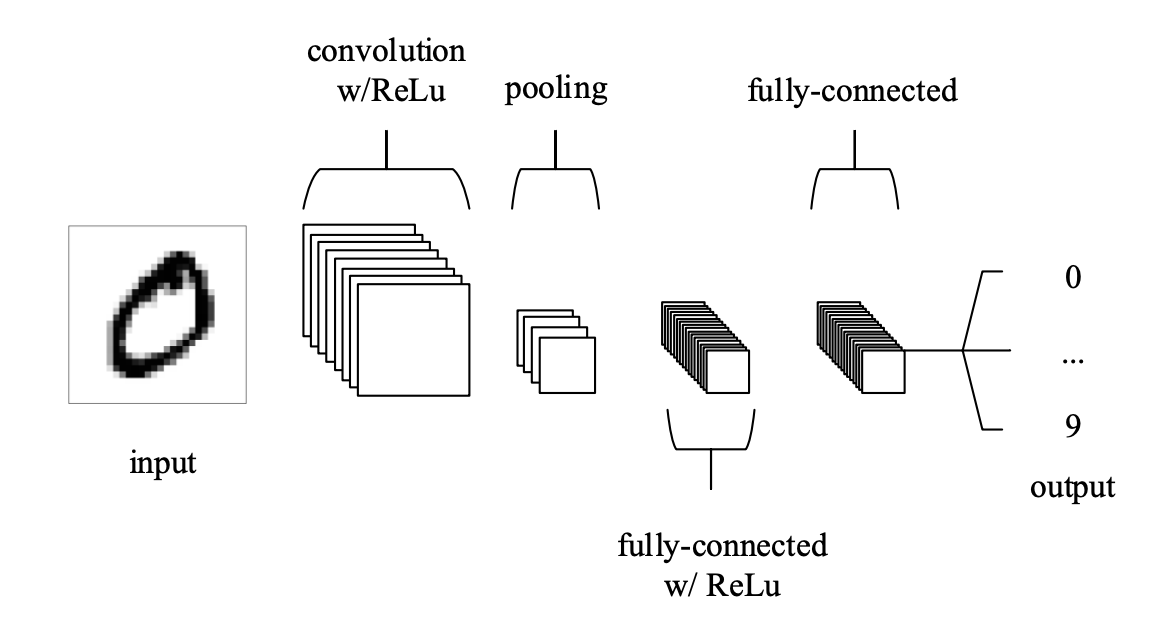}
\caption{CNN Architecture, comprised of just five layers \citep{oshea2015introduction}}
\label{cnn_architect}
\end {center}
\end{figure}

Here, input layers stores the pixel values, convolution layers determine the output of the neurons associated with local regions done by computing the scalar product between their weights and the region associated with the input volume. Rectified Linear unit (ReLu) adds the activation function to the output generated by the previous layer.

For a $z$ input value. The ReLu function is represented as: 
\begin{equation}\label{eq:relu}
    relu(\mathbf{z}) = max(0, \mathbf{z}).
\end{equation}

As per equation \ref{eq:relu},the output of ReLu is maximum between zero and input value. It can be re-written as.
 \begin{equation}\label{eq:relu_derivative}
            \nabla (relu(\mathbf{z})) =
                \begin{cases}
                \mathbf{z},  & \text{if } \mathbf{z} \geq 0 ,\\
                    0,          & \text{if } \mathbf{z} < 0 
                   
            \end{cases}
\end{equation}

The pooling layer downsamples the spatial dimensionality of the given input  and the fully-connected layers generate class values from the activations, which are used for classification.

\subsection{Convolutional layer}
Convolutional layer focus around the use of learnable kernels/filters that perform convolution operations as it is scanning the input I with respect to its dimensions giving resulting output \textbf{O} is called activation map or feature map.

\begin{figure}[H]
\begin {center}
\includegraphics[width=0.5\textwidth]{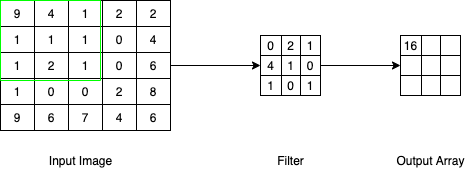}
\caption{Visual representation of a convolutional layer.}
\label{convolutionLayer}
\end {center}
\end{figure}

As seen in \autoref{convolutionLayer}, filter map value is connected to the receptive field, where  filter is being applied
The hyperparameters  that affects the resulting output size are  \textbf{Stride}, \textbf{Number of filters}, and \textbf{Zero-padding}. 
\subsubsection{Stride}
It is the distance, moved by the kernel across the input matrix. While stride values of two or more are rare, a larger stride results in a smaller output.
\subsubsection{Number of filters}
Number of filters affects the depth of the output as each individual filters yield different feature maps of its own depth. 
\subsubsection{Zero-padding}
It is mostly used when the filters do not fit the input image. In this case, all elements that fall outside the input matrix are set to zero, resulting in an output that is larger or the same size. \textit{Valid padding}, \textit{Full padding} and \textit{Same padding} are three different types of padding that can be used.

\subsection{Pooling layer}
Pooling layer aim to gradually reduce the spatial size of the representation to reduce the number of parameters and computations in the network. The pooling layer works with each feature map independently. It maps the input and scales its dimensionality with the function "MAX". The most common approach to pooling is max-pooling.
\subsection{Fully-connected layer}
Fully-connected layer contains neurons that are directly connected to neurons in the two adjacent layers, without being connected to any of the layers in them. This is consistent with the way neurons are arranged in traditional forms of ANN shown in \autoref{cnn_architect}.

\begin{figure}[H]
\begin {center}
\hspace*{-1cm}\includegraphics[width=0.57\textwidth]{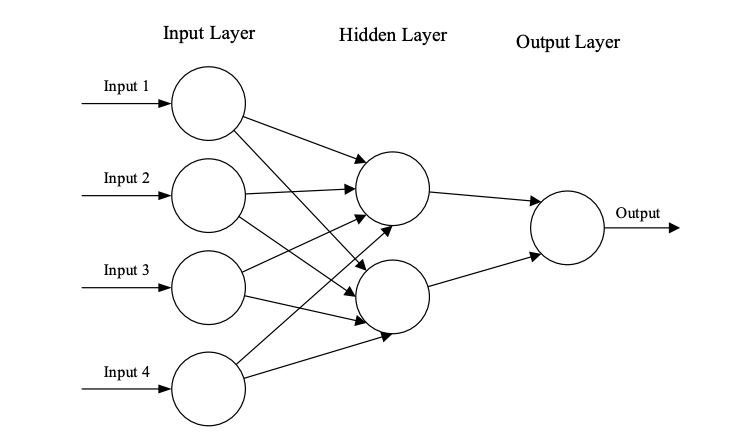}
\caption{Simple three-layer feedforward neural network (FNN) consisting of an input layer, a hidden layer, and an output layer.  \citep{oshea2015introduction}}
\label{cnn_architect}
\end {center}
\end{figure}

\section{Performance Indicators}
Different quality metrics such as Mean Square Error (MSE), Peak Signal to Noise Ratio (PSNR), Structural Similarity Index (SMI) are used to measure the quality of compressed results \citep{deshmukh_image_2019-1}. Metrics like MSE, PSNR are easy to calculate and applicable in most cases, but they sometimes do not correspond to perceived quality and are not normalized in presentation. The structured similarity indexing method (SSIM) and Feature similarity indexing method (FSIM) come into place to solve this problem\citep{sara_image_2019}.

\subsection{Mean Square Error (MSE)}
Mean Square Error(MSE) is one of the most common image error metrics used to compare image compression quality. It represents the cumulative squared error between the compressed and the original image.  The lower the MSE, the lower the error. 

The Mean Square Error can be calculated with the following expression:


 \begin{equation}\label{eq:relu_derivative}
\boxed{%
\begin{gathered}
\mathit{MSE}=\frac{1}{mn} \sum_{i=1}^m \sum_{j=1}^n (x_{ij}-y_{ij})^2 \\[1ex]
\begin{array}{@{}r|*{3}{c@{~~}}}
m      & number \ of \ rows \ in \ cover \ image\\
n      & number \ of \ columns \ in \ cover image\\
x_{ij} & pixel \ value \ from \ cover \ image\\
y_{ij} & pixel \ value \ from \ stego \ image\\
\end{array} 
\end{gathered}}
\end{equation}

\subsection{Peak Signal to Noise Ratio (PSNR)}
PSNR stands for peak signal to noise ratio. This ratio is often used as a quality measure between the original and a compressed image. In the quality degradation of image and video compression, the PSNR value varies between 30 and 50 dB for the representation of 8-bit data and between 60 and 80 dB for 16-bit data. For wireless transmission, the accepted range of quality loss is about 20 - 25 dB \citep{sara_image_2019}.
The higher the PSNR value, the better the quality of the compressed or reconstructed image.

The PSNR can be calculated with the following expression:
 \begin{equation}\label{eq:relu_derivative}
\boxed{%
\mathit{PSNR}(x,y)=\frac{10\log_{10}[\max(\max(x),\max(y))]^2}{(x-y)^2}\\
}
\end{equation}

\subsection{Structural Similarity Index (SMI)}
Structural Similarity Index is a perception-based model  that measures the structural similarity between images. Here, image degradation is considered as a change in the perception of structural information.

The SMI can be calculated with the following expression:

\begin{equation}\label{eq:relu_derivative}
\includegraphics[width=0.30\textwidth]{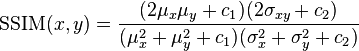}
\end{equation}

\section{Literature Survey}
 In the last decades, deep learning has been recognized as one of the most effective methods for managing large amounts of data and is widely used in the literature. Interest in more hidden layers beyond classical algorithms has recently developed in image coding and compression. In this section, previous work on the use of Deep Learning for image compression is summarized:

Experiments with two image databases, the MNIST database, and the CIFAR-10 database, show that the complexity of the learning task of NN models can be reduced when working with partially decompressed images with a small loss of accuracy of the model, while the loss of accuracy of the CNN model depends on the JPEG data. For a quality factor of 80, the gain in computational complexity with decompression is 45\% with an accuracy loss of 13\% and illustrates the need for models adapted to compressed data \citep{9105765}.

Transmitting much less data (60\%) of an image at the sender side in transform-based compression standards such as JPEG and use of a deep residual learning model on the receiver's side to recover the original data is possible with a maximum signal-to-noise ratio of over 31 dB, as in cloud servers \citep{9093951}. 

An algorithm for image compression is developed based on a back-propagation (BP) network after preprocessing the image. The proposed scheme is implemented, transfer functions influences and compression ratios within the scheme are investigated, and found out that signal-to-noise ratio (PSNR) remains almost the same for all compression ratios, while the mean square error (MSE) varies \citep{5460757}.

The relationship between pixels is highly nonlinear and unpredictable without prior knowledge of the image itself. An algorithm was created using  Artificial neural networks, that take into account the psycho-visual features  When the algorithm is applied to the image data, most of the features of the data are preserved while it is lossy and maximizes the compression performance \citep{5376811}. 

Comparison of six different image compression algorithms with a subjective quality test using high-resolution images shows that the learned compression optimized by MS-SSIM gives competitive results close to the efficiency of state-of-the-art compression \citep{8803824}.  

With the face representation model, face images are decomposed into shape and texture components, and different compression methods are applied to these components depending on the degree of redundancy. A convolutional neural network (CNN) is proposed to process the texture components. The texture components are optimized and found out that the proposed method achieved better performance in terms of PSNR and SSIM values at low bit rates in comparison to the traditional JPEG and JPEG2000 compression methods, \citep{9299680}.

The study found that when a certain generative adversarial network (GAN) is used, the compression performance of color images is better than that of gray images. To improve the compression effect of gray images, a post-processing method is proposed in this paper \citep{8958605}.

 Two recent deep-learning-based image compression algorithms are subjectively evaluated and compared with JPEG 2000 and the new BPG image codec based on HEVC Intra. The research found that compression approaches based on deep auto-encoders can achieve higher coding performance than JPEG 2000 and sometimes even as good as BPG, and suggests shows that deep generative models are likely to bring tremendous innovations in video coding in the coming years \citep{8547064}.
 
 Multilayer perceptron for image compression is presented using neural networks and used for transformation coding. Experiments are done on different images segmented into blocks of different sizes for the compression process. The reconstructed image is compared with the original image based on the signal-to-noise ratio and the number of bits per pixel. The results show that it is possible to use multilayer perceptrons for image compression \citep{4127496}.

The use of deep neural network models on embedded or mobile devices is problematic for
storage cause of modal size and large memory usage. A compression method for deep neural networks to reduce resource consumption for efficient visual inference is developed.
Four modules were introduced here: approximation module to reduce the number of weights in each linked layer, quantization module for analysis and representation of distributed weights in each layer, pruning module to suppresses small weights resulting in a reduced number of parameters, and the end coding module that represents and optimize the sparse structure of the pruned weights with relative index by Huffman coding. The proposed framework, together with the model compression and memory allocation algorithms found effective in representative models, AlexNet and VGG-16, for object recognition and face verification tasks \citep{8019465}.

An effective machine learning-based method to distinguish between double and single compressed JPEG images was introduced. Difference of JPEG 2D arrays is used to improve the compression artifacts followed by applied Markov random process to the modeling of the 2D difference arrays to utilize second-order statistics along with thresholding technique to reduce the size of the transition probability matrices. Elements of these matrices are collected as features for double JPEG compression detection and a support vector machine is used as the classifier. Experiments have shown that the proposed system outperforms the previous methods \citep{4761645}.

A reduction in the time required to train machine learning models can translate into an accuracy improvement.  For larger models, offloading of temporary data from limited GPU memory to CPU memory results in performance degradation. A new model JPEG for ACTivations (JPEGACT), was introduced that discards redundant spatial information. It adapts the well-known JPEG algorithm from 2D image compression to activation compression. This method achieves 2.4× higher training performance offload accelerators, 1.6× compression methods and consumes less than 1 percent power and area of modern CPU \citep{9138963}.

An efficient lossy image compression methods based on asymmetric autoencoders and decoder pruning was introduced through deep learning and the results show the effectiveness of the methods \citep{9053102}.

To minimize the feature distortions caused by the JPEG artifacts a new training method named Feature consistency Training was introduced. Here, in each iteration, the raw image and its compressed version of randomly sampled quality were introduced to the training model. With the addition of feature consistency constraints to the objective function,
feature distortion is minimized in the representation space to learn robust filters \citep{8933151}.

Images are compressed either to save storage space or for fast transmission requiring a time-consuming decompression step in deep modals. To address this, an architecture was purposed for object detection based on the block-wise discrete cosine transform (DCT) coefficients output from the JPEG compression algorithm in a neural network. Single-shot multibox detector (SSD) is replaced by its first layers with a convolutional layer dedicated to the processing of the DCT inputs.
Evaluations on PASCAL VOC and an industrial dataset of road traffic surveillance images show that the model is about 2× faster than the regular SSD and has promising detection performance \citep{8916937}.



\section{Discussion}

The research concludes that over the course of time there’s an advancement of deep learning in image compression. JPEG being mostly being used on the internet provides flexibility in compression with a trade-off between image quality and size. With the adaptability of neural network content, valuable information is extracted and used as a model. The use of the texture and features in neural networks improves image compression and in some cases, post-processing is also needed. The need of model adaption for a better compression using CNN is also pointed here. Most of the research shows promising results and also pointed to the need for further research.

\section{Conclusion}
JPEG is a widely used image format on the internet and provides a high compression ratio. The use of deep learning provides flexibility for experiments and also eliminates the necessity of highly complex modeling methods. It has provided promising results in the areas like image compression, image classification, image segmentation, and many more. In this review, the working mechanism of JPEG, DFT in relation to JPEG and CNN is discussed with performance indicators, different deep learning-based image compression techniques were pointed out for new studies based on JPEG compression techniques and deep learning.


%




%


\bibliography{example_bib.bib}

\begin{thebibliography}{}

\bibitem [\protect \citeauthoryear {%
Cavigelli%
, Magno%
\BCBL {}\ \BBA {} Benini%
}{%
Cavigelli%
\ \protect \BOthers {.}}{%
{\protect \APACyear {2015}}%
}]{%
cavigelli_accelerating_2015}
\APACinsertmetastar {%
cavigelli_accelerating_2015}%
\begin{APACrefauthors}%
Cavigelli, L.%
, Magno, M.%
\BCBL {}\ \BBA {} Benini, L.%
\end{APACrefauthors}%
\unskip\
\newblock
\APACrefYearMonthDay{2015}{{\APACmonth{06}}}{}.
\newblock
{\BBOQ}\APACrefatitle {Accelerating real-time embedded scene labeling with
  convolutional networks} {Accelerating real-time embedded scene labeling with
  convolutional networks}.{\BBCQ}
\newblock
\BIn{} \APACrefbtitle {Proceedings of the 52nd {Annual} {Design} {Automation}
  {Conference}} {Proceedings of the 52nd {Annual} {Design} {Automation}
  {Conference}}\ (\BPGS\ 1--6).
\newblock
\APACaddressPublisher{San Francisco California}{ACM}.
\newblock
\begin{APACrefURL}
  [{2021-09-25}]\url{https://dl.acm.org/doi/10.1145/2744769.2744788}
  \end{APACrefURL}
\newblock
\begin{APACrefDOI} \doi{10.1145/2744769.2744788} \end{APACrefDOI}
\PrintBackRefs{\CurrentBib}

\bibitem [\protect \citeauthoryear {%
Chen%
, Shi%
\BCBL {}\ \BBA {} Su%
}{%
Chen%
\ \protect \BOthers {.}}{%
{\protect \APACyear {2008}}%
}]{%
4761645}
\APACinsertmetastar {%
4761645}%
\begin{APACrefauthors}%
Chen, C.%
, Shi, Y\BPBI Q.%
\BCBL {}\ \BBA {} Su, W.%
\end{APACrefauthors}%
\unskip\
\newblock
\APACrefYearMonthDay{2008}{}{}.
\newblock
{\BBOQ}\APACrefatitle {A machine learning based scheme for double JPEG
  compression detection} {A machine learning based scheme for double jpeg
  compression detection}.{\BBCQ}
\newblock
\BIn{} \APACrefbtitle {2008 19th International Conference on Pattern
  Recognition} {2008 19th international conference on pattern recognition}\
  (\BPG~1-4).
\newblock
\begin{APACrefDOI} \doi{10.1109/ICPR.2008.4761645} \end{APACrefDOI}
\PrintBackRefs{\CurrentBib}

\bibitem [\protect \citeauthoryear {%
Cheng%
, Akyazi%
, Sun%
, Katto%
\BCBL {}\ \BBA {} Ebrahimi%
}{%
Cheng%
\ \protect \BOthers {.}}{%
{\protect \APACyear {2019}}%
}]{%
8803824}
\APACinsertmetastar {%
8803824}%
\begin{APACrefauthors}%
Cheng, Z.%
, Akyazi, P.%
, Sun, H.%
, Katto, J.%
\BCBL {}\ \BBA {} Ebrahimi, T.%
\end{APACrefauthors}%
\unskip\
\newblock
\APACrefYearMonthDay{2019}{}{}.
\newblock
{\BBOQ}\APACrefatitle {Perceptual Quality Study on Deep Learning Based Image
  Compression} {Perceptual quality study on deep learning based image
  compression}.{\BBCQ}
\newblock
\BIn{} \APACrefbtitle {2019 IEEE International Conference on Image Processing
  (ICIP)} {2019 ieee international conference on image processing (icip)}\
  (\BPG~719-723).
\newblock
\begin{APACrefDOI} \doi{10.1109/ICIP.2019.8803824} \end{APACrefDOI}
\PrintBackRefs{\CurrentBib}

\bibitem [\protect \citeauthoryear {%
Deguerre%
, Chatelain%
\BCBL {}\ \BBA {} Gasso%
}{%
Deguerre%
\ \protect \BOthers {.}}{%
{\protect \APACyear {2019}}%
}]{%
8916937}
\APACinsertmetastar {%
8916937}%
\begin{APACrefauthors}%
Deguerre, B.%
, Chatelain, C.%
\BCBL {}\ \BBA {} Gasso, G.%
\end{APACrefauthors}%
\unskip\
\newblock
\APACrefYearMonthDay{2019}{}{}.
\newblock
{\BBOQ}\APACrefatitle {Fast object detection in compressed JPEG Images} {Fast
  object detection in compressed jpeg images}.{\BBCQ}
\newblock
\BIn{} \APACrefbtitle {2019 IEEE Intelligent Transportation Systems Conference
  (ITSC)} {2019 ieee intelligent transportation systems conference (itsc)}\
  (\BPG~333-338).
\newblock
\begin{APACrefDOI} \doi{10.1109/ITSC.2019.8916937} \end{APACrefDOI}
\PrintBackRefs{\CurrentBib}

\bibitem [\protect \citeauthoryear {%
Deshmukh%
}{%
Deshmukh%
}{%
{\protect \APACyear {2019}}%
}]{%
deshmukh_image_2019-1}
\APACinsertmetastar {%
deshmukh_image_2019-1}%
\begin{APACrefauthors}%
Deshmukh, K\BPBI R.%
\end{APACrefauthors}%
\unskip\
\newblock
\APACrefYear{2019}.
\unskip\
\newblock
\APACrefbtitle {Image compression using neural networks} {Image compression
  using neural networks}\ \APACtypeAddressSchool {Master of {Science}}{San
  Jose, CA, USA}{San Jose State University}.
\unskip\
\newblock
\begin{APACrefDOI} \doi{10.31979/etd.h8mt-65ct} \end{APACrefDOI}
\PrintBackRefs{\CurrentBib}

\bibitem [\protect \citeauthoryear {%
Dong%
, Loy%
, He%
\BCBL {}\ \BBA {} Tang%
}{%
Dong%
\ \protect \BOthers {.}}{%
{\protect \APACyear {2014}}%
}]{%
10.1007/978-3-319-10593-2_13}
\APACinsertmetastar {%
10.1007/978-3-319-10593-2_13}%
\begin{APACrefauthors}%
Dong, C.%
, Loy, C\BPBI C.%
, He, K.%
\BCBL {}\ \BBA {} Tang, X.%
\end{APACrefauthors}%
\unskip\
\newblock
\APACrefYearMonthDay{2014}{}{}.
\newblock
{\BBOQ}\APACrefatitle {Learning a Deep Convolutional Network for Image
  Super-Resolution} {Learning a deep convolutional network for image
  super-resolution}.{\BBCQ}
\newblock
\BIn{} D.~Fleet, T.~Pajdla, B.~Schiele\BCBL {}\ \BBA {} T.~Tuytelaars\ (\BEDS),
  \APACrefbtitle {Computer Vision -- ECCV 2014} {Computer vision -- eccv 2014}\
  (\BPGS\ 184--199).
\newblock
\APACaddressPublisher{Cham}{Springer International Publishing}.
\PrintBackRefs{\CurrentBib}

\bibitem [\protect \citeauthoryear {%
Dosovitskiy%
\ \protect \BOthers {.}}{%
Dosovitskiy%
\ \protect \BOthers {.}}{%
{\protect \APACyear {2015}}%
}]{%
dosovitskiy_flownet:_2015}
\APACinsertmetastar {%
dosovitskiy_flownet:_2015}%
\begin{APACrefauthors}%
Dosovitskiy, A.%
, Fischer, P.%
, Ilg, E.%
, Hausser, P.%
, Hazirbas, C.%
, Golkov, V.%
\BDBL {}Brox, T.%
\end{APACrefauthors}%
\unskip\
\newblock
\APACrefYearMonthDay{2015}{{\APACmonth{12}}}{}.
\newblock
{\BBOQ}\APACrefatitle {Flownet: learning optical flow with convolutional
  networks} {Flownet: learning optical flow with convolutional
  networks}.{\BBCQ}
\newblock
\BIn{} \APACrefbtitle {2015 {IEEE} {International} {Conference} on {Computer}
  {Vision} ({ICCV})} {2015 {IEEE} {International} {Conference} on {Computer}
  {Vision} ({ICCV})}\ (\BPGS\ 2758--2766).
\newblock
\APACaddressPublisher{Santiago}{IEEE}.
\newblock
\begin{APACrefURL}
  [{2021-09-25}]\url{https://ieeexplore.ieee.org/document/7410673/}
  \end{APACrefURL}
\newblock
\begin{APACrefDOI} \doi{10.1109/ICCV.2015.316} \end{APACrefDOI}
\PrintBackRefs{\CurrentBib}

\bibitem [\protect \citeauthoryear {%
Dutta%
, Choudhury%
, Hussain%
\BCBL {}\ \BBA {} Majumder%
}{%
Dutta%
\ \protect \BOthers {.}}{%
{\protect \APACyear {2009}}%
}]{%
5376811}
\APACinsertmetastar {%
5376811}%
\begin{APACrefauthors}%
Dutta, D\BPBI P.%
, Choudhury, S\BPBI D.%
, Hussain, M\BPBI A.%
\BCBL {}\ \BBA {} Majumder, S.%
\end{APACrefauthors}%
\unskip\
\newblock
\APACrefYearMonthDay{2009}{}{}.
\newblock
{\BBOQ}\APACrefatitle {Digital Image Compression Using Neural Networks}
  {Digital image compression using neural networks}.{\BBCQ}
\newblock
\BIn{} \APACrefbtitle {2009 International Conference on Advances in Computing,
  Control, and Telecommunication Technologies} {2009 international conference
  on advances in computing, control, and telecommunication technologies}\
  (\BPG~116-120).
\newblock
\begin{APACrefDOI} \doi{10.1109/ACT.2009.38} \end{APACrefDOI}
\PrintBackRefs{\CurrentBib}

\bibitem [\protect \citeauthoryear {%
Evans%
, Liu%
\BCBL {}\ \BBA {} Aamodt%
}{%
Evans%
\ \protect \BOthers {.}}{%
{\protect \APACyear {2020}}%
}]{%
9138963}
\APACinsertmetastar {%
9138963}%
\begin{APACrefauthors}%
Evans, R\BPBI D.%
, Liu, L.%
\BCBL {}\ \BBA {} Aamodt, T\BPBI M.%
\end{APACrefauthors}%
\unskip\
\newblock
\APACrefYearMonthDay{2020}{}{}.
\newblock
{\BBOQ}\APACrefatitle {JPEG-ACT: Accelerating Deep Learning via Transform-based
  Lossy Compression} {Jpeg-act: Accelerating deep learning via transform-based
  lossy compression}.{\BBCQ}
\newblock
\BIn{} \APACrefbtitle {2020 ACM/IEEE 47th Annual International Symposium on
  Computer Architecture (ISCA)} {2020 acm/ieee 47th annual international
  symposium on computer architecture (isca)}\ (\BPG~860-873).
\newblock
\begin{APACrefDOI} \doi{10.1109/ISCA45697.2020.00075} \end{APACrefDOI}
\PrintBackRefs{\CurrentBib}

\bibitem [\protect \citeauthoryear {%
Ge%
, Luo%
, Zhao%
, Jin%
\BCBL {}\ \BBA {} Zhang%
}{%
Ge%
\ \protect \BOthers {.}}{%
{\protect \APACyear {2017}}%
}]{%
8019465}
\APACinsertmetastar {%
8019465}%
\begin{APACrefauthors}%
Ge, S.%
, Luo, Z.%
, Zhao, S.%
, Jin, X.%
\BCBL {}\ \BBA {} Zhang, X\BHBI Y.%
\end{APACrefauthors}%
\unskip\
\newblock
\APACrefYearMonthDay{2017}{}{}.
\newblock
{\BBOQ}\APACrefatitle {Compressing deep neural networks for efficient visual
  inference} {Compressing deep neural networks for efficient visual
  inference}.{\BBCQ}
\newblock
\BIn{} \APACrefbtitle {2017 IEEE International Conference on Multimedia and
  Expo (ICME)} {2017 ieee international conference on multimedia and expo
  (icme)}\ (\BPG~667-672).
\newblock
\begin{APACrefDOI} \doi{10.1109/ICME.2017.8019465} \end{APACrefDOI}
\PrintBackRefs{\CurrentBib}

\bibitem [\protect \citeauthoryear {%
Hu%
\ \protect \BOthers {.}}{%
Hu%
\ \protect \BOthers {.}}{%
{\protect \APACyear {2020}}%
}]{%
9299680}
\APACinsertmetastar {%
9299680}%
\begin{APACrefauthors}%
Hu, S.%
, Duan, Y.%
, Tao, X.%
, Liu, Y.%
, Zhang, X.%
\BCBL {}\ \BBA {} Lu, J.%
\end{APACrefauthors}%
\unskip\
\newblock
\APACrefYearMonthDay{2020}{}{}.
\newblock
{\BBOQ}\APACrefatitle {Content-aware Facial Image Compression with Deep
  Learning Method} {Content-aware facial image compression with deep learning
  method}.{\BBCQ}
\newblock
\BIn{} \APACrefbtitle {2020 International Conference on Wireless Communications
  and Signal Processing (WCSP)} {2020 international conference on wireless
  communications and signal processing (wcsp)}\ (\BPG~516-521).
\newblock
\begin{APACrefDOI} \doi{10.1109/WCSP49889.2020.9299680} \end{APACrefDOI}
\PrintBackRefs{\CurrentBib}

\bibitem [\protect \citeauthoryear {%
Ifeachor%
\ \BBA {} Jervis%
}{%
Ifeachor%
\ \BBA {} Jervis%
}{%
{\protect \APACyear {2002}}%
}]{%
ifeachor_digital_2002}
\APACinsertmetastar {%
ifeachor_digital_2002}%
\begin{APACrefauthors}%
Ifeachor, E\BPBI C.%
\BCBT {}\ \BBA {} Jervis, B\BPBI W.%
\end{APACrefauthors}%
\unskip\
\newblock
\APACrefYear{2002}.
\newblock
\APACrefbtitle {Digital signal processing: a practical approach} {Digital
  signal processing: a practical approach}\ (\PrintOrdinal{2nd ed}\ \BEd).
\newblock
\APACaddressPublisher{Harlow, England ; New York}{Prentice Hall}.
\PrintBackRefs{\CurrentBib}

\bibitem [\protect \citeauthoryear {%
Kim%
, Choi%
, Chang%
\BCBL {}\ \BBA {} Lee%
}{%
Kim%
\ \protect \BOthers {.}}{%
{\protect \APACyear {2020}}%
}]{%
9053102}
\APACinsertmetastar {%
9053102}%
\begin{APACrefauthors}%
Kim, J\BHBI H.%
, Choi, J\BHBI H.%
, Chang, J.%
\BCBL {}\ \BBA {} Lee, J\BHBI S.%
\end{APACrefauthors}%
\unskip\
\newblock
\APACrefYearMonthDay{2020}{}{}.
\newblock
{\BBOQ}\APACrefatitle {Efficient Deep Learning-Based Lossy Image Compression
  Via Asymmetric Autoencoder and Pruning} {Efficient deep learning-based lossy
  image compression via asymmetric autoencoder and pruning}.{\BBCQ}
\newblock
\BIn{} \APACrefbtitle {ICASSP 2020 - 2020 IEEE International Conference on
  Acoustics, Speech and Signal Processing (ICASSP)} {Icassp 2020 - 2020 ieee
  international conference on acoustics, speech and signal processing
  (icassp)}\ (\BPG~2063-2067).
\newblock
\begin{APACrefDOI} \doi{10.1109/ICASSP40776.2020.9053102} \end{APACrefDOI}
\PrintBackRefs{\CurrentBib}

\bibitem [\protect \citeauthoryear {%
Krizhevsky%
, Sutskever%
\BCBL {}\ \BBA {} Hinton%
}{%
Krizhevsky%
\ \protect \BOthers {.}}{%
{\protect \APACyear {2017}}%
}]{%
krizhevsky_imagenet_2017}
\APACinsertmetastar {%
krizhevsky_imagenet_2017}%
\begin{APACrefauthors}%
Krizhevsky, A.%
, Sutskever, I.%
\BCBL {}\ \BBA {} Hinton, G\BPBI E.%
\end{APACrefauthors}%
\unskip\
\newblock
\APACrefYearMonthDay{2017}{{\APACmonth{05}}}{}.
\newblock
{\BBOQ}\APACrefatitle {{ImageNet} classification with deep convolutional neural
  networks} {{ImageNet} classification with deep convolutional neural
  networks}.{\BBCQ}
\newblock
\APACjournalVolNumPages{Communications of the ACM}{60}{6}{84--90}.
\newblock
\begin{APACrefURL} [{2021-10-06}]\url{https://dl.acm.org/doi/10.1145/3065386}
  \end{APACrefURL}
\newblock
\begin{APACrefDOI} \doi{10.1145/3065386} \end{APACrefDOI}
\PrintBackRefs{\CurrentBib}

\bibitem [\protect \citeauthoryear {%
Kunwar%
}{%
Kunwar%
}{%
{\protect \APACyear {2017}}%
}]{%
kunwar_image_2017}
\APACinsertmetastar {%
kunwar_image_2017}%
\begin{APACrefauthors}%
Kunwar, S.%
\end{APACrefauthors}%
\unskip\
\newblock
\APACrefYearMonthDay{2017}{}{}.
\newblock
{\BBOQ}\APACrefatitle {Image {Compression} {Algorithm} and {JPEG} {Standard}}
  {Image {Compression} {Algorithm} and {JPEG} {Standard}}.{\BBCQ}
\newblock
\APACjournalVolNumPages{International Journal of Scientific and Research
  Publications}{7}{12}{150--157}.
\newblock
\begin{APACrefURL}
  \url{http://www.ijsrp.org/research-paper-1217/ijsrp-p7224.pdf}
  \end{APACrefURL}
\PrintBackRefs{\CurrentBib}

\bibitem [\protect \citeauthoryear {%
Kunwar%
}{%
Kunwar%
}{%
{\protect \APACyear {2018}}%
}]{%
kunwar_jpeg_2018}
\APACinsertmetastar {%
kunwar_jpeg_2018}%
\begin{APACrefauthors}%
Kunwar, S.%
\end{APACrefauthors}%
\unskip\
\newblock
\APACrefYearMonthDay{2018}{{\APACmonth{01}}}{}.
\newblock
{\BBOQ}\APACrefatitle {Jpeg image compression using cnn} {Jpeg image
  compression using cnn}.{\BBCQ}
\newblock
\APACjournalVolNumPages{}{}{}{1}.
\newblock
\begin{APACrefURL}
  [{2021-09-26}]\url{http://rgdoi.net/10.13140/RG.2.2.25600.53762}
  \end{APACrefURL}
\newblock
\begin{APACrefDOI} \doi{10.13140/RG.2.2.25600.53762} \end{APACrefDOI}
\PrintBackRefs{\CurrentBib}

\bibitem [\protect \citeauthoryear {%
Lim%
\ \BBA {} Oppenheim%
}{%
Lim%
\ \BBA {} Oppenheim%
}{%
{\protect \APACyear {1988}}%
}]{%
lim_advanced_1988}
\APACinsertmetastar {%
lim_advanced_1988}%
\begin{APACrefauthors}%
Lim, J., S%
\BCBT {}\ \BBA {} Oppenheim, A\BPBI V.%
\end{APACrefauthors}%
\unskip\
\newblock
\APACrefYear{1988}.
\newblock
\APACrefbtitle {Advanced topics in signal processing} {Advanced topics in
  signal processing}.
\newblock
\APACaddressPublisher{London}{Prentice-Hall International}.
\newblock
\APACrefnote{OCLC: 857116358}
\PrintBackRefs{\CurrentBib}

\bibitem [\protect \citeauthoryear {%
Long%
, Shelhamer%
\BCBL {}\ \BBA {} Darrell%
}{%
Long%
\ \protect \BOthers {.}}{%
{\protect \APACyear {2015}}%
}]{%
7298965}
\APACinsertmetastar {%
7298965}%
\begin{APACrefauthors}%
Long, J.%
, Shelhamer, E.%
\BCBL {}\ \BBA {} Darrell, T.%
\end{APACrefauthors}%
\unskip\
\newblock
\APACrefYearMonthDay{2015}{}{}.
\newblock
{\BBOQ}\APACrefatitle {Fully convolutional networks for semantic segmentation}
  {Fully convolutional networks for semantic segmentation}.{\BBCQ}
\newblock
\BIn{} \APACrefbtitle {2015 IEEE Conference on Computer Vision and Pattern
  Recognition (CVPR)} {2015 ieee conference on computer vision and pattern
  recognition (cvpr)}\ (\BPG~3431-3440).
\newblock
\begin{APACrefDOI} \doi{10.1109/CVPR.2015.7298965} \end{APACrefDOI}
\PrintBackRefs{\CurrentBib}

\bibitem [\protect \citeauthoryear {%
Mitra%
}{%
Mitra%
}{%
{\protect \APACyear {2006}}%
}]{%
mitra_digital_2006}
\APACinsertmetastar {%
mitra_digital_2006}%
\begin{APACrefauthors}%
Mitra, S\BPBI K.%
\end{APACrefauthors}%
\unskip\
\newblock
\APACrefYear{2006}.
\newblock
\APACrefbtitle {Digital signal processing: a computer based approach} {Digital
  signal processing: a computer based approach}\ (\PrintOrdinal{3rd ed}\ \BEd).
\newblock
\APACaddressPublisher{New York, NY}{McGraw-Hill Higher Education}.
\newblock
\APACrefnote{OCLC: ocm56011040}
\PrintBackRefs{\CurrentBib}

\bibitem [\protect \citeauthoryear {%
O'Shea%
\ \BBA {} Nash%
}{%
O'Shea%
\ \BBA {} Nash%
}{%
{\protect \APACyear {2015}}%
}]{%
oshea2015introduction}
\APACinsertmetastar {%
oshea2015introduction}%
\begin{APACrefauthors}%
O'Shea, K.%
\BCBT {}\ \BBA {} Nash, R.%
\end{APACrefauthors}%
\unskip\
\newblock
\APACrefYearMonthDay{2015}{}{}.
\newblock
\APACrefbtitle {An Introduction to Convolutional Neural Networks.} {An
  introduction to convolutional neural networks.}
\PrintBackRefs{\CurrentBib}

\bibitem [\protect \citeauthoryear {%
Pistono%
, Coatrieux%
, Nunes%
\BCBL {}\ \BBA {} Cozic%
}{%
Pistono%
\ \protect \BOthers {.}}{%
{\protect \APACyear {2020}}%
}]{%
9105765}
\APACinsertmetastar {%
9105765}%
\begin{APACrefauthors}%
Pistono, M.%
, Coatrieux, G.%
, Nunes, J\BHBI C.%
\BCBL {}\ \BBA {} Cozic, M.%
\end{APACrefauthors}%
\unskip\
\newblock
\APACrefYearMonthDay{2020}{}{}.
\newblock
{\BBOQ}\APACrefatitle {Training Machine Learning on JPEG Compressed Images}
  {Training machine learning on jpeg compressed images}.{\BBCQ}
\newblock
\BIn{} \APACrefbtitle {2020 Data Compression Conference (DCC)} {2020 data
  compression conference (dcc)}\ (\BPG~388-388).
\newblock
\begin{APACrefDOI} \doi{10.1109/DCC47342.2020.00070} \end{APACrefDOI}
\PrintBackRefs{\CurrentBib}

\bibitem [\protect \citeauthoryear {%
Poor%
}{%
Poor%
}{%
{\protect \APACyear {1988}}%
}]{%
poor_introduction_1988}
\APACinsertmetastar {%
poor_introduction_1988}%
\begin{APACrefauthors}%
Poor, H\BPBI V.%
\end{APACrefauthors}%
\unskip\
\newblock
\APACrefYear{1988}.
\newblock
\APACrefbtitle {An introduction to signal detection and estimation} {An
  introduction to signal detection and estimation}\ (J\BPBI B.~Thomas, \BED{}).
\newblock
\APACaddressPublisher{New York, NY}{Springer New York}.
\newblock
\begin{APACrefURL}
  [{2021-09-25}]\url{http://link.springer.com/10.1007/978-1-4757-3863-6}
  \end{APACrefURL}
\newblock
\begin{APACrefDOI} \doi{10.1007/978-1-4757-3863-6} \end{APACrefDOI}
\PrintBackRefs{\CurrentBib}

\bibitem [\protect \citeauthoryear {%
Pratt%
}{%
Pratt%
}{%
{\protect \APACyear {2006}}%
}]{%
pratt2006a}
\APACinsertmetastar {%
pratt2006a}%
\begin{APACrefauthors}%
Pratt, W.%
\end{APACrefauthors}%
\unskip\
\newblock
\APACrefYear{2006}.
\newblock
\APACrefbtitle {Digital Image Processing} {Digital image processing}.
\newblock
\APACaddressPublisher{New York, NY, USA}{John Wiley \& Sons, Ltd}.
\PrintBackRefs{\CurrentBib}

\bibitem [\protect \citeauthoryear {%
Qiu%
\ \protect \BOthers {.}}{%
Qiu%
\ \protect \BOthers {.}}{%
{\protect \APACyear {2021}}%
}]{%
9093951}
\APACinsertmetastar {%
9093951}%
\begin{APACrefauthors}%
Qiu, H.%
, Zheng, Q.%
, Memmi, G.%
, Lu, J.%
, Qiu, M.%
\BCBL {}\ \BBA {} Thuraisingham, B.%
\end{APACrefauthors}%
\unskip\
\newblock
\APACrefYearMonthDay{2021}{}{}.
\newblock
{\BBOQ}\APACrefatitle {Deep Residual Learning-Based Enhanced JPEG Compression
  in the Internet of Things} {Deep residual learning-based enhanced jpeg
  compression in the internet of things}.{\BBCQ}
\newblock
\APACjournalVolNumPages{IEEE Transactions on Industrial
  Informatics}{17}{3}{2124-2133}.
\newblock
\begin{APACrefDOI} \doi{10.1109/TII.2020.2994743} \end{APACrefDOI}
\PrintBackRefs{\CurrentBib}

\bibitem [\protect \citeauthoryear {%
K\BPBI R.~Rao%
\ \BBA {} Ochoa-Dominguez%
}{%
K\BPBI R.~Rao%
\ \BBA {} Ochoa-Dominguez%
}{%
{\protect \APACyear {2019}}%
}]{%
rao_discrete_2019}
\APACinsertmetastar {%
rao_discrete_2019}%
\begin{APACrefauthors}%
Rao, K\BPBI R.%
\BCBT {}\ \BBA {} Ochoa-Dominguez, H.%
\end{APACrefauthors}%
\unskip\
\newblock
\APACrefYear{2019}.
\newblock
\APACrefbtitle {Discrete cosine transform} {Discrete cosine transform}\
  (\PrintOrdinal{Second edition}\ \BEd).
\newblock
\APACaddressPublisher{Boca Raton, FL}{Taylor \& Francis Group, CRC Press}.
\PrintBackRefs{\CurrentBib}

\bibitem [\protect \citeauthoryear {%
P\BPBI V.~Rao%
, Madhusudana%
, Nachiketh%
\BCBL {}\ \BBA {} Keerthi%
}{%
P\BPBI V.~Rao%
\ \protect \BOthers {.}}{%
{\protect \APACyear {2010}}%
}]{%
5460757}
\APACinsertmetastar {%
5460757}%
\begin{APACrefauthors}%
Rao, P\BPBI V.%
, Madhusudana, S.%
, Nachiketh, S.%
\BCBL {}\ \BBA {} Keerthi, K.%
\end{APACrefauthors}%
\unskip\
\newblock
\APACrefYearMonthDay{2010}{}{}.
\newblock
{\BBOQ}\APACrefatitle {Image Compression using Artificial Neural Networks}
  {Image compression using artificial neural networks}.{\BBCQ}
\newblock
\BIn{} \APACrefbtitle {2010 Second International Conference on Machine Learning
  and Computing} {2010 second international conference on machine learning and
  computing}\ (\BPG~121-124).
\newblock
\begin{APACrefDOI} \doi{10.1109/ICMLC.2010.33} \end{APACrefDOI}
\PrintBackRefs{\CurrentBib}

\bibitem [\protect \citeauthoryear {%
Ren%
, He%
, Girshick%
\BCBL {}\ \BBA {} Sun%
}{%
Ren%
\ \protect \BOthers {.}}{%
{\protect \APACyear {2016}}%
}]{%
ren_faster_2016}
\APACinsertmetastar {%
ren_faster_2016}%
\begin{APACrefauthors}%
Ren, S.%
, He, K.%
, Girshick, R.%
\BCBL {}\ \BBA {} Sun, J.%
\end{APACrefauthors}%
\unskip\
\newblock
\APACrefYearMonthDay{2016}{{\APACmonth{01}}}{}.
\newblock
{\BBOQ}\APACrefatitle {Faster r-cnn: towards real-time object detection with
  region proposal networks} {Faster r-cnn: towards real-time object detection
  with region proposal networks}.{\BBCQ}
\newblock
\APACjournalVolNumPages{arXiv:1506.01497 [cs]}{}{}{}.
\newblock
\begin{APACrefURL} [{2021-09-25}]\url{http://arxiv.org/abs/1506.01497}
  \end{APACrefURL}
\newblock
\APACrefnote{arXiv: 1506.01497}
\PrintBackRefs{\CurrentBib}

\bibitem [\protect \citeauthoryear {%
Ruihua%
, Quan%
\BCBL {}\ \BBA {} Huachao%
}{%
Ruihua%
\ \protect \BOthers {.}}{%
{\protect \APACyear {2019}}%
}]{%
8958605}
\APACinsertmetastar {%
8958605}%
\begin{APACrefauthors}%
Ruihua, L.%
, Quan, Z.%
\BCBL {}\ \BBA {} Huachao, X.%
\end{APACrefauthors}%
\unskip\
\newblock
\APACrefYearMonthDay{2019}{}{}.
\newblock
{\BBOQ}\APACrefatitle {An Image Compression Processing Method Based On Deep
  Learning} {An image compression processing method based on deep
  learning}.{\BBCQ}
\newblock
\BIn{} \APACrefbtitle {2019 IEEE 2nd International Conference on Information
  Communication and Signal Processing (ICICSP)} {2019 ieee 2nd international
  conference on information communication and signal processing (icicsp)}\
  (\BPG~342-346).
\newblock
\begin{APACrefDOI} \doi{10.1109/ICICSP48821.2019.8958605} \end{APACrefDOI}
\PrintBackRefs{\CurrentBib}

\bibitem [\protect \citeauthoryear {%
Sara%
, Akter%
\BCBL {}\ \BBA {} Uddin%
}{%
Sara%
\ \protect \BOthers {.}}{%
{\protect \APACyear {2019}}%
}]{%
sara_image_2019}
\APACinsertmetastar {%
sara_image_2019}%
\begin{APACrefauthors}%
Sara, U.%
, Akter, M.%
\BCBL {}\ \BBA {} Uddin, M\BPBI S.%
\end{APACrefauthors}%
\unskip\
\newblock
\APACrefYearMonthDay{2019}{}{}.
\newblock
{\BBOQ}\APACrefatitle {Image quality assessment through fsim, ssim, mse and
  psnr—a comparative study} {Image quality assessment through fsim, ssim, mse
  and psnr—a comparative study}.{\BBCQ}
\newblock
\APACjournalVolNumPages{Journal of Computer and Communications}{07}{03}{8--18}.
\newblock
\begin{APACrefURL}
  [{2021-10-04}]\url{http://www.scirp.org/journal/doi.aspx?DOI=10.4236/jcc.2019.73002}
  \end{APACrefURL}
\newblock
\begin{APACrefDOI} \doi{10.4236/jcc.2019.73002} \end{APACrefDOI}
\PrintBackRefs{\CurrentBib}

\bibitem [\protect \citeauthoryear {%
Valenzise%
, Purica%
, Hulusic%
\BCBL {}\ \BBA {} Cagnazzo%
}{%
Valenzise%
\ \protect \BOthers {.}}{%
{\protect \APACyear {2018}}%
}]{%
8547064}
\APACinsertmetastar {%
8547064}%
\begin{APACrefauthors}%
Valenzise, G.%
, Purica, A.%
, Hulusic, V.%
\BCBL {}\ \BBA {} Cagnazzo, M.%
\end{APACrefauthors}%
\unskip\
\newblock
\APACrefYearMonthDay{2018}{}{}.
\newblock
{\BBOQ}\APACrefatitle {Quality Assessment of Deep-Learning-Based Image
  Compression} {Quality assessment of deep-learning-based image
  compression}.{\BBCQ}
\newblock
\BIn{} \APACrefbtitle {2018 IEEE 20th International Workshop on Multimedia
  Signal Processing (MMSP)} {2018 ieee 20th international workshop on
  multimedia signal processing (mmsp)}\ (\BPG~1-6).
\newblock
\begin{APACrefDOI} \doi{10.1109/MMSP.2018.8547064} \end{APACrefDOI}
\PrintBackRefs{\CurrentBib}

\bibitem [\protect \citeauthoryear {%
Vilovic%
}{%
Vilovic%
}{%
{\protect \APACyear {2006}}%
}]{%
4127496}
\APACinsertmetastar {%
4127496}%
\begin{APACrefauthors}%
Vilovic, I.%
\end{APACrefauthors}%
\unskip\
\newblock
\APACrefYearMonthDay{2006}{}{}.
\newblock
{\BBOQ}\APACrefatitle {An Experience in Image Compression Using Neural
  Networks} {An experience in image compression using neural networks}.{\BBCQ}
\newblock
\BIn{} \APACrefbtitle {Proceedings ELMAR 2006} {Proceedings elmar 2006}\
  (\BPG~95-98).
\newblock
\begin{APACrefDOI} \doi{10.1109/ELMAR.2006.329523} \end{APACrefDOI}
\PrintBackRefs{\CurrentBib}

\bibitem [\protect \citeauthoryear {%
Wallace%
}{%
Wallace%
}{%
{\protect \APACyear {1991}}%
}]{%
wallace_jpeg_1991}
\APACinsertmetastar {%
wallace_jpeg_1991}%
\begin{APACrefauthors}%
Wallace, G\BPBI K.%
\end{APACrefauthors}%
\unskip\
\newblock
\APACrefYearMonthDay{1991}{{\APACmonth{04}}}{}.
\newblock
{\BBOQ}\APACrefatitle {The {JPEG} still picture compression standard} {The
  {JPEG} still picture compression standard}.{\BBCQ}
\newblock
\APACjournalVolNumPages{Communications of the ACM}{34}{4}{30--44}.
\newblock
\begin{APACrefURL}
  [{2021-10-04}]\url{https://dl.acm.org/doi/10.1145/103085.103089}
  \end{APACrefURL}
\newblock
\begin{APACrefDOI} \doi{10.1145/103085.103089} \end{APACrefDOI}
\PrintBackRefs{\CurrentBib}

\bibitem [\protect \citeauthoryear {%
Wan%
, Wu%
, Hsu%
, Wong%
\BCBL {}\ \BBA {} Lee%
}{%
Wan%
\ \protect \BOthers {.}}{%
{\protect \APACyear {2020}}%
}]{%
8933151}
\APACinsertmetastar {%
8933151}%
\begin{APACrefauthors}%
Wan, S.%
, Wu, T\BHBI Y.%
, Hsu, H\BHBI W.%
, Wong, W\BPBI H.%
\BCBL {}\ \BBA {} Lee, C\BHBI Y.%
\end{APACrefauthors}%
\unskip\
\newblock
\APACrefYearMonthDay{2020}{}{}.
\newblock
{\BBOQ}\APACrefatitle {Feature Consistency Training With JPEG Compressed
  Images} {Feature consistency training with jpeg compressed images}.{\BBCQ}
\newblock
\APACjournalVolNumPages{IEEE Transactions on Circuits and Systems for Video
  Technology}{30}{12}{4769-4780}.
\newblock
\begin{APACrefDOI} \doi{10.1109/TCSVT.2019.2959815} \end{APACrefDOI}
\PrintBackRefs{\CurrentBib}

\bibitem [\protect \citeauthoryear {%
Wang%
, Bovik%
\BCBL {}\ \BBA {} Lu%
}{%
Wang%
\ \protect \BOthers {.}}{%
{\protect \APACyear {2021}}%
}]{%
wang2021a}
\APACinsertmetastar {%
wang2021a}%
\begin{APACrefauthors}%
Wang, Z.%
, Bovik, A.%
\BCBL {}\ \BBA {} Lu, L.%
\end{APACrefauthors}%
\unskip\
\newblock
\APACrefYearMonthDay{2021}{}{}.
\newblock
\APACrefbtitle {WHY IS IMAGE QUALITY ASSESSMENT SO DIFFICULT?} {Why is image
  quality assessment so difficult?}
\newblock
\begin{APACrefURL}
  \url{https://live.ece.utexas.edu/publications/2002/zw_icassp2002_whyqa.pdf}
  \end{APACrefURL}
\newblock
\APACrefnote{online] Live.ece.utexas.edu. Available at:}
\PrintBackRefs{\CurrentBib}

\bibitem [\protect \citeauthoryear {%
Zhang%
, Cai%
\BCBL {}\ \BBA {} Xiong%
}{%
Zhang%
\ \protect \BOthers {.}}{%
{\protect \APACyear {2021}}%
}]{%
9088290}
\APACinsertmetastar {%
9088290}%
\begin{APACrefauthors}%
Zhang, Y.%
, Cai, Z.%
\BCBL {}\ \BBA {} Xiong, G.%
\end{APACrefauthors}%
\unskip\
\newblock
\APACrefYearMonthDay{2021}{}{}.
\newblock
{\BBOQ}\APACrefatitle {A New Image Compression Algorithm Based on Non-Uniform
  Partition and U-System} {A new image compression algorithm based on
  non-uniform partition and u-system}.{\BBCQ}
\newblock
\APACjournalVolNumPages{IEEE Transactions on Multimedia}{23}{}{1069-1082}.
\newblock
\begin{APACrefDOI} \doi{10.1109/TMM.2020.2992940} \end{APACrefDOI}
\PrintBackRefs{\CurrentBib}

\end{thebibliography}



    
  





\newpage
\onecolumn












\end{document}